\documentclass[aps,10pt,twocolumn,groupeaddress,showpacs,reprintnumbers,floatfix]{revtex4}
\usepackage{epsfig}
\usepackage{longtable}
\usepackage{graphics}
\usepackage{graphicx}
\usepackage{lscape}
\usepackage{dcolumn}
\begin{document}
\title{Quasi-Particle and Collective Magnetism: Rotation, Pairing and Blocking in High-K Isomers}
\author{N.~J.~Stone}
\affiliation{Department of Physics, University of Oxford, Oxford, OX1 3PU, UK}
\affiliation {Department of Physics and Astronomy, University of Tennessee, Knoxville, TN 37996, USA} 
\author{J.~R.~Stone}
\affiliation{Department of Physics, University of Oxford, Oxford, OX1 3PU, UK}
\affiliation {Department of Physics and Astronomy, University of Tennessee, Knoxville, TN 37996, USA} 
\author{P.~M.~Walker}
\affiliation{Department of Physics, University of Surrey, Guildford, Surrey, GU2 7XH, UK}
\author{C.~R.~Bingham}
\affiliation {Department of Physics and Astronomy, University of Tennessee, Knoxville, TN 37996, USA} 
\affiliation {Physics Division, Oak Ridge National Laboratory, Oak Ridge, TN 37831, USA}
\date{\today}
 \begin{abstract}
For the first time, a wide range of collective magnetic g-factors g$_{\rm R}$, obtained from a novel analysis of experimental
 data for multi-quasiparticle configurations in high-K isomers, is shown to exhibit a striking systematic variation with the relative number of proton
 and neutron quasiparticles,  N$_{\rm p}$ - N$_{\rm n}$.  Using the principle of additivity, the quasi-particle contribution to magnetism in high-K isomers
 of Lu - Re, Z = 71 - 75, has been estimated. Based on these estimates, band-structure branching ratio data are used to explore the behaviour
 of the collective contribution as the number and  proton/neutron nature (N$_{\rm p}$, N$_{\rm n}$), of the quasi-particle
 excitations, change. Basic ideas of pairing, its quenching by quasi-particle excitation and the consequent
 changes to moment of inertia and collective magnetism are discussed. Existing model calculations do not reproduce the observed  g$_{\rm R}$ variation adequately. The paired superfluid system of nucleons in these nuclei, and their excitations, present properties of general physics interest. The new-found systematic behaviour
 of g$_{\rm R}$ in multi-quasi-particle excitations of this unique system, showing variation from close to zero for multi-neutron states to above 0.5 for multi-proton states, opens a fresh window on these effects and raises the important question of just which nucleons contribute to the `collective' properties of these nuclei.
\end{abstract}
\maketitle
The motion of charges in physical systems gives rise to magnetic fields that themselves provide key observables for understanding the underlining structures - ranging from plasma clouds in spiral galaxies to superfluid flow in rotating atomic nuclei. In the latter case both neutrons and protons carry angular momentum, but it is mainly the protons that provide the magnetism. Furthermore, the single-particle contributions of unpaired nucleons (quasi-particles) in deformed nuclei need to be distinguished from the core of collectively rotating paired nucleons. The present work focuses on the separation between these single-particle and collective components, shedding light on the breakdown of nuclear superfluidity by quasi-particle (qp) excitations.

Nuclei of elements between Z = 71 (Lu) and Z = 75 (Re) show strong, approximately constant, deformation and have been well described by the axially symmetric deformation model. In this model qp motion is described in terms of a quantum number $K$, the projection of the qp angular momentum {\bf{J}} on the axis of deformation and collective motion is  associated with quantum number $R$ which describes rotation about an axis perpendicular to the deformation axis. The spin of each nuclear state is the vector sum, {\bf{I}} = {\bf{J}} + {\bf{R}}. Nuclei in the region typically exhibit rotational bands built upon each qp state which forms the band head. The band heads are either single-qp states in the deformed potential, or multi-qp states made up of combinations of single-qp states. Frequently such multiple states are isomeric, having total K values differing from neighbouring states. These states, known as high-K isomers, which have well defined qp make-up, and the bands built upon them, have been the subject of intense investigation since the first was found in the early 1950's \cite{hf1953}.  Both qp and collective components of the state wavefunction contribute to the magnetic moment of the states, through g-factors g$_{\rm K}$ and g$_{\rm R}$, the total moment being given by the expression (for general state of spin I in a band built upon a band-head of intrinsic spin K)
\begin{equation}
\mu = g_{\rm R}I + (g_{\rm K} - g_{\rm R})\frac{K^{\rm 2}}{I + 1}
\label{eq1}
\end{equation}	
which for the bandhead, I = K, becomes
\begin{equation}
\mu = g_{\rm K} [I^{\rm 2}/(I + 1)]  + g_{\rm R}[I/(I + 1)]
\label{eq2}
\end{equation}
and for band states built on the ground state (K = 0) of even-even nuclei
\begin{equation}
\mu = g_{\rm R} I
\label{eq3}
\end{equation}

Two properties of the band states, the E2/M1 multipole mixing ratio $\delta$ in transitions between states of a band having spins differing by one, and the branching ratio $\lambda$ in the decay of a band state of spin I to lower states in the band having spin I - 1 and spin I - 2 also depend upon both g$_{\rm K}$ and g$_{\rm R}$, through the expressions \cite{purry1998}
\begin{equation}
\frac{\delta^{\rm 2}}{1+\delta^{\rm 2}}=\frac{2K^{\rm 2}(2I-1)}{(I+1)(I-1+K)(I-1-K)} \frac{E^{\rm 5}_{\rm \gamma,\Delta I=1}}{E^{\rm 5}_{\rm \gamma,\Delta I=2}} \lambda
\label{eq4}
\end{equation}
and, with E$_\gamma$ in MeV and Q$_{\rm 0}$, the intrinsic quadrupole moment, in eb,
\begin{equation}
\frac{g_{\rm K} - g_{\rm R}}{Q_{\rm 0}}=\frac{0.933E_{\rm \gamma,\Delta I=1}}{\delta \sqrt{I^{\rm 2}-1}}
\label{eq5}
\end{equation}
						     		   	
It has been a widespread feature of analysis of high-K isomers that their g$_{\rm K}$ factors are used to assist identification of the qp make-up of the state. To this end, empirical g$_{\rm K_{\rm i}}$ values of individual qps of spin K$_{\rm i}$ are adopted based on magnetic moment measurements on single-qp band head (I = K$_{\rm i}$) states of similar deformation in the same region of nuclei. The adopted empirical g$_{\rm K_{\rm i}}$ values then give estimates for g$_{\rm K}$ of multi-qp states of total spin K = $\sum{K_{\rm i}}$ using the relation
\begin{equation}
g_{\rm K} = \frac{1}{K}\sum_{i}K_{\rm i}g_{\rm K_{\rm i}}.
\label{add}
\end{equation}
 
There are few measurements of magnetic moments of high-K isomers and relatively few of the required E2/M1 mixing ratios so that usually the estimated g$_{\rm K}$ values are used, with assumed values of g$_{\rm R}$ and estimates of Q$_{\rm 0}$, to compare with measured branching ratios $\lambda$ as an aid to identification of the qp configuration. The values taken for g$_{\rm R}$ have been set within a small range, between 0.25 and 0.35. No variation in g$_{\rm R}$ has been considered as the number or identity of the qps involved in the state under consideration changes within a single isotope.

Thus to date, although the valuable physics information contained in the parameter g$_{\rm R}$ has been recognized (see e.g.\cite{mullins2000, purry1998,nilsson1964}), in the field of high K-isomers no systematic investigation has been made. Classically g$_{\rm R}$ relates directly to the degree to which the rotating body carries charge, so that, for a nucleus made up of Z protons and N neutrons rotating as a uniform solid body, the simple expectation is g$_{\rm R}$ = Z/(Z + N) or Z/A. However the fact that nuclear rotation was more complex was recognized from the 1960's and early studies of rotation bands in even-even nuclei yielded moments of inertia (MoI) which were much reduced from classical body predictions. These reductions were understood as being associated with the phenomenon of pairing, whereby nucleon pairs, treated as bosons, fell into a superfluid state and did not contribute to the MoI. The number of nucleons involved in the superfluid state is a function of the pairing strength. As the rotational frequency increases in higher members of the band, nucleon pairs are broken and the MoI rises towards the classical value \cite{nilsson1964,bm}. 

In relation to g$_{\rm R}$ factors we note that pairing affects protons and neutrons separately. The effect of breaking a pair, as in odd-A nuclei, should block that state from participation in the pairing effect and so weaken the effect of pairing for the nucleon type involved. Thus pairing affects the contributions of protons and neutrons to the MoI and to g$_{\rm R}$. Bohr and Mottelson \cite{bm} in their landmark 1975 text remark that g$_{\rm R}$ values for single-qp states having a broken proton/neutron pair should be changed compared to the even-even neighbour of the same element. They tabulate evidence, based on data then available, that odd proton single-qp states have higher g$_{\rm R}$ values and odd neutron single-particle states lower g$_{\rm R}$ values than their even-even neighbours (see \cite{bm} Table 5-14, Vol II). Further direct discussion of the variation in g$_{\rm R}$ in this region has been very limited, despite the great activity in study of high-K isomers in the intervening years.
\begin{table*}
\caption{\label{states}Adopted g$_{\rm K_{\rm i}}$ value for individual qp states. Experimental moments taken from \cite{stone} with g$_{\rm R}$ = 0.29(5).} 
\vspace{5pt} 
\begin{tabular}{ccll}
\hline
qp  &       Adopted g$_{\rm K_{\rm i}}$     & Basis of adopted value\\  \hline
Protons         &                 &      \\    \hline
5/2+[402]       &       1.67(6)   &              moments of 5/2+ 482 keV state in $^{\rm 181}$Ta  \\
                &                 &              and ground states of $^{\rm 181,183,185,187}$Re\\
7/2-[523]       &       1.04(5)   &              moment of 5/2- 379 keV state in $^{\rm 169}$Tm \\
7/2+[404]       &       0.765(25) &              moments of 7/2+ ground states in $^{\rm 175,177,179,181}$Ta  \\
9/2-[514]       &       1.37(3)   &              moment of 9/2- 6 keV state in $^{\rm 181}$Ta  \\
11/2-[505]      &       1.281(14) &              moment of 11/2- 434 keV state in $^{\rm 187}$Ir \\  \hline
Neutrons        &                 &     \\   \hline
5/2-[512]       &       -0.48(2)        &       moment of 5/2- ground state of $^{\rm 175}$Hf  \\
7/2-[514]       &       0.206(14)       &       moment of 7/2- ground state of $^{\rm 177}$Hf  \\
7/2-[503]       &       -0.319(15)      &       moment of 7/2- ground state of $^{\rm 157}$Yb  \\
7/2+[633]       &       -0.323(18)      &       moment of 7/2+ ground state of $^{\rm 175}$W   \\
9/2+[624]       &       -0.239(11)      &       moment of 9/2+ ground state of $^{\rm 179}$Hf   \\  \hline
\end{tabular}
\end{table*}
 
The present investigation was prompted by the results of recent experimental work on isomers of $^{\rm 177}$Hf carried out at the NICOLE on-line nuclear orientation facility at ISOLDE, CERN \cite{bingham2011}. These and earlier results gave accurate g$_{\rm R}$ factors for bands based on states 7/2- (0 keV), g$_{\rm R}$  = 0.24(2), 9/2+ (321 KeV), g$_{\rm R}$ = 0.21(2), 23/2+ (1335 keV), g$_{\rm R}$ = 0.30(3) and 37/2- (2790 keV), g$_{\rm R}$ = 0.21(4) to be considered with the neighbour $^{\rm 178}$Hf where the qp vacuum 2$^{\rm +}_{\rm 1}$ state has g$_{\rm R}$ = 0.280(7), the 6+ state (1554 keV) has g$_{\rm R}$ = 0.43(6) and the 16+ state (2446 keV) has g$_{\rm R}$= 0.27(2) (for configurations see Supplementary Material Table SMII). The marked g$_{\rm R}$ variation in two adjacent nuclei demanded broader study.

In the analysis reported in this letter a new approach has been adopted. High-K isomers, involving states having single, multiple and combinations of quasi-neutron and quasi-proton excitations, frequently exist in the same nucleus and certainly in close neighbours with very similar nuclear deformation. Thus, if good estimates can be made of the g$_{\rm K}$ parameters of the band heads in these nuclei, g$_{\rm R}$ values can be extracted from branching ratios and the influence of specific qp excitations on pairing and the superfluid state explored with a new perspective.

To establish the potential of this approach it is necessary to test the quality of estimates of the g$_{\rm K}$ parameter. For this we have taken the assumption of additivity as expressed in relation (\ref{add}).  Comparison is made between estimates based on the adopted g$_{\rm K_{\rm i}}$ values and experimental g-factors measured in multi-qp states in the selected region. The chosen g$_{\rm K_{\rm i}}$ values are given in Table~\ref{states}, with the basis for their choice. All are derived from measured moments of states of single-qp states in the region. Table~\ref{config} compares the estimates derived from additivity based on these g$_{\rm Ki}$ values with experimental measurements. Although making this comparison requires use of an estimate of g$_{\rm R}$ (see Eq.~\ref{eq2}), which could give rise to a circular argument, an important feature of expressions (\ref{eq2}) and (\ref{eq5}) allows the procedure to be followed without problems. The weight of the g$_{\rm R}$ term in expression (\ref{eq2}) is less, by a factor $K$ of the state, than the weight of the g$_{\rm K}$ term, whereas, in (\ref{eq5}) the g$_{\rm R}$ and g$_{\rm K}$ terms have equal weight. Since $K$ has values of six and above in this study, an approximate value of g$_{\rm R}$ can be used in (\ref{eq2}) to estimate the moment and give rise to only a small additional uncertainty, never more than a few \%, in the resulting moment prediction. This is shown in detail in Supplementary Material Table SMI.

\begin{table*}
\scriptsize
\squeezetable
\caption{\label{config}Check for additivity in magnetic moments of multi-qp states. g$_{\rm K}$ taken from Table~\ref{states} with g$_{\rm R}$ = 0.29(5) using Eq.~\ref{eq2}. Experimental moment values are from \protect\cite{stone}.}
\vspace{5pt} 
\begin{tabular}{ccccccccc}
\hline																
Z  &   Isotope & I$^{\rm \pi}$ & \multicolumn{2}{c}{qps} & Kg$_{\rm K}$ & \multicolumn{2}{c}{ $\mu$ [$\mu_{\rm N}$]}	& (est/exp - 1)\%  \\  \hline
   &          &      &     protons  &	neutrons	&           &  est   &  exp &   \\  \hline 															
76 &	$^{\rm 182}$Os	& 25$^+$  &	9/2-[514] 11/2-[505]   &   7/2-[503] 7/2-[514]   &   10.56(19) &	10.43(19) &	10.62(20)	& -2(3)\\
   &                     &          &                           &   7/2+[633] 9/2+[624]   &             &               &         &      \\	
75 &	$^{\rm 182}$Re	  & 16$^+$   &	9/2-[514] 	           &   9/2+[624] 7/2+[633] 7/2-[514]		  &   4.68(17)  &	4.69(16)  &	 3.82(13)	& 23(4)\\																
															
75 &	$^{\rm 182}$Re	  & 16$^+$   &	9/2-[514] 	           &   9/2+[624] 7/2-[503] 7/2+[633]		  &   2.88(17)  &	2.98(17)  &	 3.82(13)	& -22(7)\\																
75 &	$^{\rm 182}$Re	  & 16$^-$   &	9/2-[514]	           &   9/2+[624] 7/2-[514] 7/2-[503]		  &   4.72(15)  &  4.72(15) &  3.82(13)   &  24(4) \\
																		
72 &	$^{\rm 177}$Hf	  & 37/2$-$ &   7/2+[404] 9/2-[514]    &   5/2-[512] 7/2-[514] 9/2+[624]		  &   7.29(18)  &	7.20(18)  &	7.33(9)    &  -2(3) \\
																		
71 &	$^{\rm 177}$Lu	  & 23/2$^+$ &	7/2+[404]	           &   7/2-[514] 9/2+[624]	                    &	2.32(11) &	2.40(11)  &	2.32(10)   &  3(6)  \\
																		
74 &	$^{\rm 179}$W       & 35/2$-$  &   5/2+[402] 9/2-[514]    &   5/2-[512] 7/2-[514] 9/2+[624]		  &	8.79(22) &	8.58(22)  &	7.5(15)    &  14(20) \\  
																		
74 &	$^{\rm 176}$W	  & 14$^+$   &	7/2+[404] 9/2-[514]    &   7/2+[633] 5/2-[512]                      &	6.51(18) & 	6.35(18)  &	6.7(2)     &  -5(4) \\
																		
72 &	$^{\rm 178}$Hf	  & 16$^+$  &   7/2+[404] 9/2-[514]    &   7/2-[514] 9/2+[624]                      &	8.49(18) &	8.26(17)  &	8.16(5)   &	 1(2) \\
																		
75 &	$^{\rm 182}$Re      & 7+     &   5/2+[402]              &   9/2+[624]					  &   3.09(16) &	2.96(15)  &	2.79(6)   &  6(6) \\
																		
72 &	$^{\rm 179}$Hf	  & 25/2$-$ &   7/2+[404] 9/2-[514]    &   9/2+[624]					  &   7.68(17) & 	7.46(17)  &	7.43(34)  &	 0(5) \\
																		
72 &	$^{\rm 178}$Hf	  & 6$^+$   &  7/2+[404] 5/2+[402]     &							  &   6.85(16) &	6.12(15)  &	5.84(5)   &  5(3) \\
																		
72 &	$^{\rm 180}$Hf	  & 8$-$     &  7/2+[404] 9/2-[514]     &							  &   8.84(16) &	8.12(15)  &	8.6(10)   &  -6(12) \\
																		
72 &	$^{\rm 172}$Hf      & 8$-$     &  7/2+[404] 9/2-[514]     &							  &   8.84(16) &	8.12(15)  &	7.96(7)   &  2(3) \\
																		
72 &	$^{\rm 172}$Hf      & 6$^+$    &  7/2+[404] 5/2+[402]     &							  &   6.85(16) &	6.12(15)  &	5.5(6)   &	11(11) \\
																		
72 &	$^{\rm 174}$Hf      &  6$^+$   &  7/2+[404] 5/2+[402]     &							  &   6.85(16) &	6.12(15)  &	5.35(4)  & 14(3)  \\
																		
72 &	$^{\rm 176}$Hf      & 6$^+$    &  7/2+[404] 5/2+[402]	    &							  &   6.85(16) &	6.12(15)  &	5.75(5)  &	6(3) \\  \hline
\end{tabular}
\end{table*}
 
\begin{figure}
\includegraphics[width=9cm]{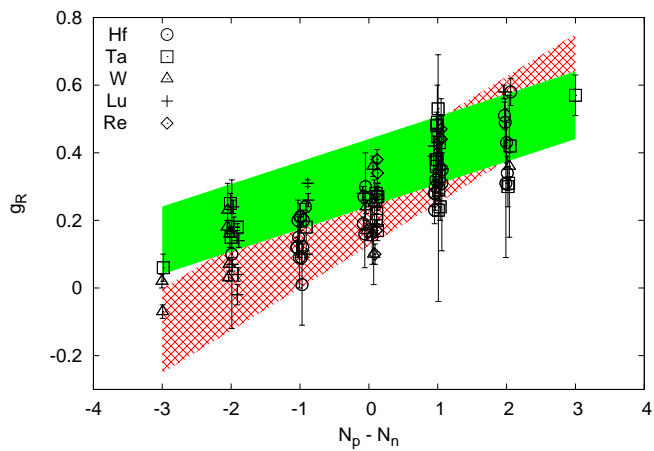}
\caption{\label{grplot}(Color on-line) Plot of extracted g$_{\rm R}$ as a function N$_{\rm p}$-N$_{\rm n}$, the difference between numbers of proton and neutron qps. In this and Fig.~\ref{grplot_th} the x coordinate is slightly displaced to add visibility. For details of bands see text.}
\end{figure}

Table~\ref{config} shows that the assumption of additivity produces magnetic moment values which deviate from experiment typically by less than 5\%. For the remainder of this work we have used the adopted g$_{\rm K_{\rm i}}$ values from Table~\ref{states} in all estimates of g$_{\rm K}$. We note that, whilst similar tests of additivity have been made before (see e.g. \cite{walker1994} Fig. 10) no attempt to use this finding to explore the behavior of g$_{\rm R}$, as reported here, has been made.
 
Having established a reliable method for estimating g$_{\rm K}$ for multi-qp states we can use published spectroscopic values of the parameter $|$(g$_{\rm K}$ - g$_{\rm R}$)/Q$_{\rm 0}$$|$, which are available from branching ratio measurements in many bands built on high-K isomeric states, to obtain values of g$_{\rm R}$ (see Supplementary  Material Table SMII for details). For the intrisic quadrupole moment Q$_{\rm 0}$ we have adopted the value 7.2 eb (see e.g.\cite{kondev1997}) with the exception of $^{\rm 185,187}$Re ground states for which the measured, lower, values have been used \cite{hagn1981}. Other values in the literature vary by, at most, 5\% from this value for all states considered, so giving a relatively small contribution to the final uncertainty in g$_{\rm R}$. The necessary choice of sign of this parameter has been made in most cases to eliminate either unrealistically high positive or clearly negative values of g$_{\rm R}$. The results are plotted in Fig.~\ref{grplot}. In a few cases the choice of sign cannot be made and both are entered in Fig.~\ref{grplot}. The errors in g$_{\rm R}$ are a compound of uncertainty in  g$_{\rm K}$ and in the branching ratio. 
\begin{figure}
\includegraphics[width=8cm]{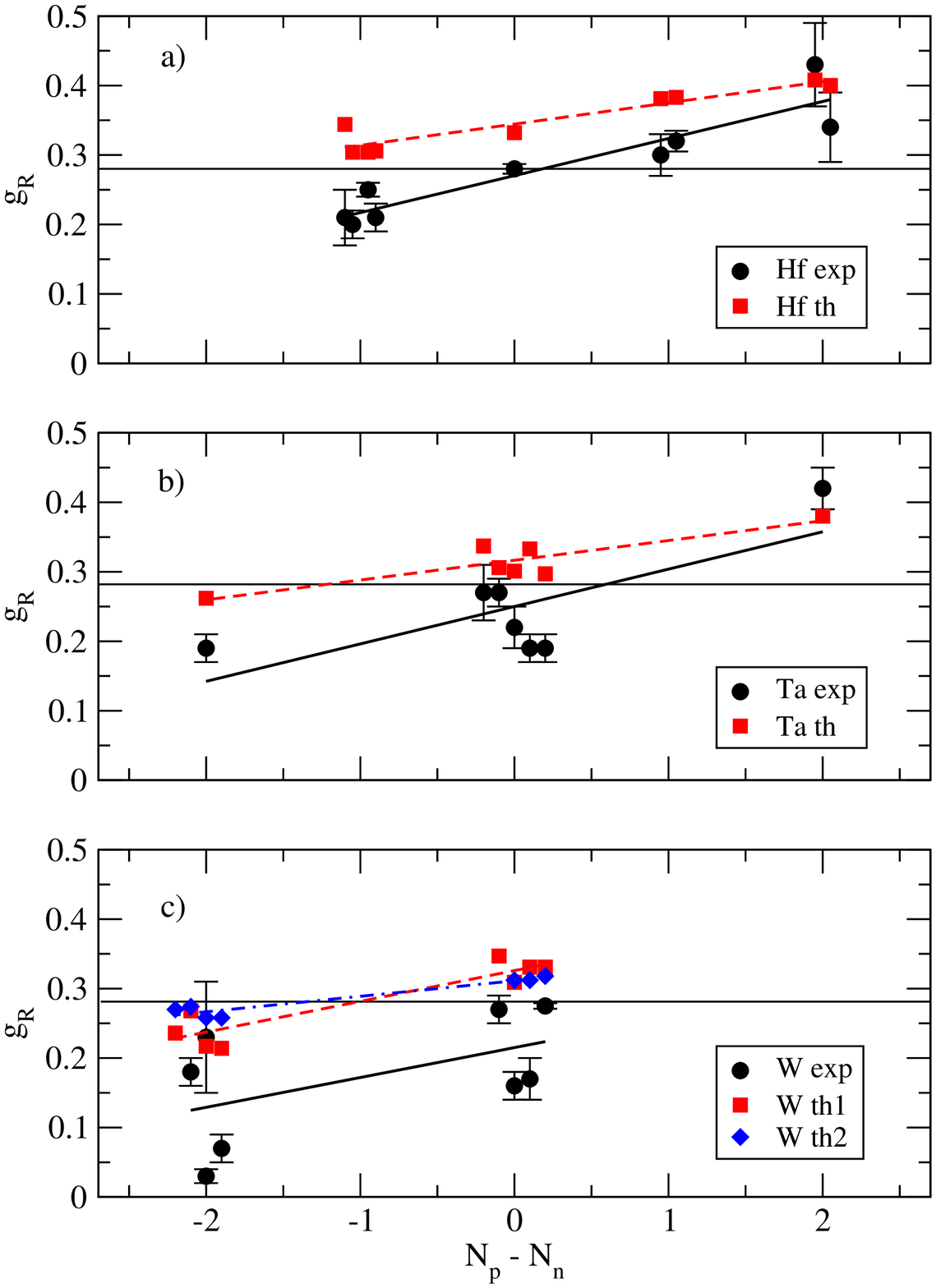}
\caption{\label{grplot_th} (Color on-line) gR obtained from experiment (this work) compared with model calculations: Hf \cite{xu2013}, Ta \cite{dracoulis1998}, W (th1) \cite{purry1998} and W (th2) \cite{dracoulis1998}.}
\end{figure}
In Fig.~\ref{grplot} the results for g$_{\rm R}$ are grouped and plotted as a function of the variable N$_{\rm p}$ - N$_{\rm n}$, the difference between the numbers of proton and neutron qps involved in the isomer. This variable was chosen since, to first order, the effects of breaking proton and neutron pairs are expected to cancel. Fig.~\ref{grplot} shows a clear systematic behavior in g$_{\rm R}$, with variation from close to zero to well above 0.5, the higher values being clearly above Z/A (of order 0.41 see Table~\ref{gr2}). As expected the lowest values are found for the greatest excess in numbers of broken neutron pairs and the highest for the most broken proton pairs. The fact that, for example at N$_{\rm p}$ - N$_{\rm n}$ = 0, there is considerable variation in g$_{\rm R}$, is understood since we do not expect exact cancellation between the effects of breaking different proton and neutron pairs. Nevertheless, the overall behaviour is striking and it represents the principal finding of the present study. For the first time, a wide range of g$_{\rm R}$ values, obtained from a novel analysis of experimental data for multi-qp configurations, is clearly seen to exhibit a simple dependence on the relative number of proton and neutron qps.

It is not the objective of this letter to offer a quantitative analysis of the variation of g$_{\rm R}$ which has been revealed by the present approach. Rather we offer the results, and some comparisons with existing empirical and theoretical observations. The variation of the g$_{\rm R}$ parameter is a challenging new and very direct window into the collective behavior of nucleons in these isotopes, in particular how pairing and superfluidity are modified by the blocking effects of specific multiple proton and neutron excitations.
\begin{table*}
\caption{\label{gr2} g-factors of first 2+ qp-vacuum states in even-even isotopes of Yb, Hf, W, and Os \cite{stone}.} 
\vspace{5pt} 
\begin{tabular}{ccc>{\centering}p{2cm}p{2cm}r}
\hline
Z	&Isotope	      &	Z/A	  &	\multicolumn{2}{c}{g$_{\rm R}$}  & average g$_{\rm R}$  \\ \hline
70	&$^{\rm 164}$Yb	&	0.427	  &	0.325(50) &           &             \\
	&$^{\rm 170} $Yb	&	0.412	  &	0.337(4)  &           &  0.338(3)    \\
	&$^{\rm 174}$Yb	&	0.402	  &	0.338(4)  &           &           \\
	&$^{\rm 176}$Yb	&	0.398	  &	0.340(15) &           &           \\
72	&$^{\rm 176}$Hf	&	0.409	  &	0.315(30) & 0.270(20) &           \\
	&$^{\rm 178}$Hf	&	0.404	  &  	0.300(20) & 0.240(15) &	 0.280(7) \\
      &$^{\rm 180}$Hf	&	0.400	  &	0.305(15) & 0.255(40) &           \\  
      &                 &             &	0.265(15) & 0.385(35) &           \\        
74	&$^{\rm 180}$W	&     0.411	  &	0.255(15) &           &           \\
	&$^{\rm 182}$W    &     0.407	  &	0.260(10) &  	    &   0.275(4) \\
	&$^{\rm 184}$W	&     0.402	  &	0.269(7)  & 0.268(7)  &            \\
	&$^{\rm 186}$W	&     0.398	  &	0.310(15) & 0.310(10) &            \\          
76	&$^{\rm 186}$Os   &     0.409	  &	0.280(10) & 0.265(10) &            \\
	&$^{\rm 188}$Os   &     0.404	  &	0.290(10) & 0.305(15) &	 0.286(5) \\
      &                 &             &   0.308(10) &           &             \\  \hline
\end{tabular}
\end{table*}

The first point of discussion takes the basic assumption that the relation \cite{nilsson1964,bm} g$_{\rm R}$ = $\mathcal{I}$$_{\rm p}$/($\mathcal{I}$$_{\rm p}$ + $\mathcal{I}$$_{\rm n}$), where $\mathcal{I}$$_{\rm p,n}$ is the proton, neutron contribution to the total moment of inertia (MoI), holds good.  We examine the change found in g$_{\rm R}$ in states having a single-qp excitation as compared to the qp vacuum in ground state bands of even-even nuclei. Differentiation of g$_{\rm R}$ with respect to change in $\mathcal{I}$$_{\rm p}$ and $\mathcal{I}$$_{\rm n}$ yields the results
\begin{equation}
\Delta g_{\rm R} = \frac{\Delta \mathcal{I}_{\rm p}(1 - g_{\rm R(e-e)})}{\mathcal{I}_{\rm p} + \mathcal{I}_{\rm n}} \qquad\mbox{for a proton excitation}\qquad
\label{eq7}	
\end{equation}	
and
\begin{equation}
\Delta g_{\rm R} = g_{\rm R(e-e)}\frac{-\Delta \mathcal{I}_{\rm n}}{\mathcal{I}_{\rm p} + \mathcal{I}_{\rm n}} \qquad\mbox{for a  neutron excitation}\qquad
\label{eq8}	
\end{equation}	
where $\Delta\mathcal{I}_{\rm p}$ and $\Delta\mathcal{I}_{\rm n}$ are the changes produced in the MoI in the single-qp nuclei as compared to their even-even neighbours and g$_{\rm R(e-e)}$ is the $2^+_{\rm 1}$ state g-factor in the neighboring even-even (qp vacuum) nuclei. Relevant $\mathcal{I}$, $\Delta\mathcal{I}$ values are found from Table 5.17 of \cite{bm}, values of g$_{\rm R(e-e)}$ are gathered in Table~\ref{gr2} and the comparison between the predicted changes in g$_{\rm R}$ and our findings is given in Table~\ref{dgr}. There is good order of magnitude agreement for both single proton and single-neutron excitation, with quantitative agreement for specific single-proton excitations.

\begin{table*}
\caption{\label{dgr}Comparison between changes in g$_{\rm R}$, estimated from Eqs. \ref{eq7} and \ref{eq8} and \cite{bm} (column 2) and experiment (this work) for single-qp bands. * depends on choice of sign of [g$_{\rm K}$ - g$_{\rm R}$] } 
\vspace{5pt} 
\begin{tabular}{lccl}
\hline
              &  $\Delta g_{\rm R}$(MoI)    &   $\Delta g_{\rm R}$ (Bands)   &   Bands analyzed in   \\   \hline
Odd Protons   &                &                      &                            \\
7/2+[404]	  &    +0.07	 &      +0.09(1)	      &     $^{\rm 175,177}$Lu,$^{\rm 175,177,179}$Ta      \\
5/2+[402]	  &    +0.09	 &      +0.10(4)	      &     $^{\rm 175,177,179}$Ta,$^{\rm 185,187}$Re         \\
9/2-[514]	  &    +0.13	 &      +0.17(2)	      &     $^{\rm 175,177,179}$Ta      \\
Odd Neutrons  &                &                      &                             \\
5/2-[512]	  &     -0.04	 &     -0.16(3)		&     $^{\rm 173,175,179}$Hf   \\
7/2-[514]	  &     -0.03	 &     -0.16(2)* or -0.03(1)		&     $^{\rm 177,179}$Hf      \\
7/2+[633]	  &     -0.07	 &     -0.18(2)		&     $^{\rm 173,175}$Hf      \\
9/2+[624]	  &     -0.09	 &     -0.075(15)	      &     $^{\rm 177}$Hf          \\    \hline
\end{tabular}
\end{table*}		
 We can take this point a stage further by exploring the degree to which g$_{\rm R}$ shifts in multi-qp states can be seen as additive. Since superconductive pairing is a collective effect it might be expected that simple additivity would not work well as the number of available pairs falls, reducing the pairing strength. Also, although we expect the combination of proton and neutron excitations to be generally canceling, the fact that the single-qp shifts of specific excitations differ should lead to a scatter in combined shifts for band-heads having different configurations but the same value of N$_{\rm p}$  $-$ N$_{\rm n}$. Nevertheless, assuming addivity we have used the single-qp shifts in Table~\ref{dgr} to predict shifts in all multi-qp bands. This has been done taking both the MoI and the band analysis single-qp shift sets. Details are given in the Supporting Material [URL] Table SMII. (The choice of shift for the 7/2-[514] neutron excitation, although it changes some individual data points, has little overall effect on the situation. The value -0.16 was taken in the results shown). The predictions are shown in Fig.~\ref{grplot} as bands, the full color based on the MoI set and the cross-hatched band on the band analyses set and shows noteworthy features. In particular, the agreement in both the slope and the variation/scatter at each each N$_{\rm p}$ $-$ N$_{\rm n}$ value are very similar to those found in the experimentally based multi-qp data in Fig.~\ref{grplot}.
There is little sign of weakening of the effect of more qp excitations that might have been expected.
The second point of discussion compares predictions for g$_{\rm R}$ derived from existing analyses of qp excitation energies with the experimental g$_{\rm R}$ values here extracted from the bands built upon them. Several models have been used to provide values of the pairing gaps for calculation of the MoI for high-K isomers in Hf, Ta and W isotopes, e.g. blocked BCS \cite{purry1998}  and Lipkin-Nogami \cite{dracoulis1998,xu2013}. Using $\Delta_{\rm p,n}$ from these analyses and the expression
\begin{equation}
\mathcal{I}_{\rm p,n}=\mathcal{I}\left(1-\frac{\ln[x_{\rm p,n}+(1+x^{\rm 2}_{\rm p,n})^{\rm 1/2}]}{x_{\rm p,n}(1+x^{\rm 2}_{p,n})^{\rm 1/2}}\right)
\end{equation}
where $\mathcal{I}$ is the rigid body value and $x_{\rm p,n}=\frac{(\delta \hbar \omega_{\rm 0})_{\rm p,n}}{2 \Delta_{\rm p,n}}$, values of $\mathcal{I}_{\rm p}$ and $\mathcal{I}_{\rm n}$ and predictions for g$_{\rm R}$ can be obtained. Comparison with experimental g$_{\rm R}$ values produced by the present analysis in the same Hf, Ta and W isomers is shown in Fig~\ref{grplot_th}. In each panel of the figure the horizontal line is at the qp vacuum g$_{\rm R}$ value. The full and dashed sloping lines are least squares fits to the respective data to guide the eye and emphasize the differences between them. Although the trend of the g$_{\rm R}$ variation with N$_{\rm p}$ $-$ N$_{\rm n}$ is correct, detailed agreement is seen to be poor with the theoretical prediction showing too little slope.  
 
To summarise, this work has used the assumption of additivity in estimation of the intrinsic qp contribution to magnetism in high-K isomers to extract information regarding the behavior of the collective contribution parameter g$_{\rm R}$. A striking systematic behavior of g$_{\rm R}$ has been found, showing wider variation that previously appreciated, for larger excess of either proton or neutron qps. The behavior is closely connected to the question of pairing and the superfluid state of protons and neutrons in nuclei, offering a sensitive new window on this phenomenon previously discussed primarily in association with reduced moments of inertia.

Simple ideas have been shown to give broadly satisfactory qualitative consistency between effects found in the g$_{\rm R}$ parameter and in the moment of inertia for single-qp excitation. Only in recent years has the availability of more extensive multi-qp excitation data allowed this variation to be extended as presented here and the more extensive and systematic variation of g$_{\rm R}$ to be revealed.

We reiterate the unique interest in this particular system, nucleons in nuclei, which exhibits properties related to pairing and superfluidity. Unlike e.g. electrons in metals and liquid He, the participating pairs in this system are few and their excitations, that is the breaking of different specific quantum states, can be expected to have different effects upon the paired system. The g$_{\rm R}$ parameter, explored here more fully than before, presents a relatively unexamined feature of this phenomenon. 

It is hoped that this work will stimulate new activity in several directions. It is clear that in many cases better precision in experimental spectroscopic studies would improve our knowledge of the variation of g$_{\rm R}$. More and precise direct magnetic moment measurements of high-K isomers would be of great value. However it would appear that existing data should be enough to encourage more theoretical work to elucidate an improved description of the collective aspects of these nuclei. \\

We thank F.~R.~Xu for performing additional Lipkin-Nogami pairing gap calculations in Hf isomers. The research was supported by US DOE Office of Science and STFC (UK).\\

Supplementary material related to this article is at\\ 
http://dx.doi.org/10.1016/j.physletb.2013.09.016


\begin{thebibliography}{99}
\bibitem{hf1953}
A.~Bohr and B.~Mottelson, Phys. Rev. 90 (1953) 717
\bibitem{mullins2000}
S.~M.~Mullins et al., Phys. Rev. C61  (2000) 044315
\bibitem{purry1998}
C.~S.~Purry et al., Nucl. Phys. A632 (1998) 229
\bibitem{nilsson1964}
S.~G.~Nilsson in \textit{Perturbed Angular Correlations}, eds. E.~Karlsson, E.~Matthias and K.~Siegbahn, North Holland, 1964, pp. 163 - 181.
\bibitem{bm}
A. Bohr and B. Mottelson, \textit{Nuclear Structure}, W. A. Benjamins, Inc, 1975
\bibitem{bingham2011}
Briefly reported at the ARIS-11 meeting, Leuven 2011 C.R.Bingham et al, and to be published.
\bibitem{walker1994}
P.~M.~Walker et al., Nucl. Phys.  A568 (1994) 397
\bibitem{kondev1997}
F.~G.~Kondev et al., Nucl. Phys. A617  (1997) 91
\bibitem{hagn1981}
E.~Hagn et al., Nucl. Phys. A363 (1981) 269
\bibitem{stone}
N. J. Stone, Table of Nuclear Magnetic Dipole and Electric quadruple Moments, IAEA Nuclear Data Section Report INDC(NDS)-0594, April 2011
\bibitem{xu2013}
F.~R.~Xu, private communication  (2013)
\bibitem{dracoulis1998}
G.~D.~Dracoulis et al., Phys. Letts. B419  (1998) 7
\end{thebibliography}
\end{document}